\documentclass[12pt,thmsa]{article}

\def \eq {\begin{equation}}
\def \fim-eq {\end{equation}}
\begin{document}

\author{E. S. Guerra \\
Departamento de F\'{\i}sica \\
Universidade Federal Rural do Rio de Janeiro \\
Cx. Postal 23851, 23890-000 Serop\'edica, RJ, Brazil \\
email: emerson@ufrrj.br\\
}
\title{REALIZATION OF GHZ STATES AND THE GHZ TEST VIA CAVITY QED FOR A
CAVITY PREPARED IN A SUPERPOSITION OF ZERO AND ONE FOCK STATES}
\maketitle

\begin{abstract}
\noindent In this article we discuss the realization of atomic GHZ states
involving three-level atoms in a cascade and in a lambda configuration and
we show explicitly how to use this state to perform the GHZ test in which it
is possible to decide between local realism theories and quantum mechanics.
The experimental realizations proposed makes use the interaction of Rydberg
atoms with a cavity prepared in a state which is a superposition of zero and
one Fock states.

\ \newline

\end{abstract}

\section{\protect\bigskip INTRODUCTION\protect\bigskip}

The superposition principle, entanglement and its consequence non-locality
makes quantum mechanics fascinating and, at the time, intriguing and not
intuitive according to our common sense which is based essentially on
classical mechanics. \ Quantum mechanics has given rise to many questions of
philosophical nature. For instance, the concept of reality \cite{JammerPhil,
Omnes, Hughes, Krips, Espagnat} has new interpretations under the light of
quantum physics. The Consequences of entanglement and non-locality were
noticed by Einstein, Podolsky and Rosen who proposed a \textit{gedanken }%
experiment, the EPR experiment in a very interesting article \cite{EPR,
Omnes, Hughes, Krips, WallsMilburn} to show that quantum theory was not a
complete theory. In their article they describe an experimental arrangement
involving correlated pair of particles where these particles interact and
then are separated. .Therefore, measurements made on one particle can be
used via correlation to generate predictions about the other particle. EPR
state that\ "If, without in any way disturbing a system we can predict with
certainty the value of a physical quantity, then there exists an element of
physical reality corresponding to this physical quantity" and as a necessary
condition for the completeness of the theory, EPR state that \ "every
element of the physical reality must have a counter part in the physical
theory". EPR noticed that the fact that quantum mechanics admits no
dispersion-free states, does not tell us wether the theory is complete or
not. In their words: \ "From [the dispersion principle] it follows (1) the
quantum mechanics description of reality given by the wave function is not
complete or (2) when operators corresponding to two physical quantities do
not commute the two quantities cannot have simultaneous reality". Shortly
after the EPR paper Bohr published a reply in which he defends the
completeness of the quantum-mechanical description of nature which could not
be refuted properly by EPR\ and which came to be known as the Copenhagen
interpretation \cite{BohrEPRreply}.

Hidden\ variable theories \cite{Omnes, Hughes} were \ proposed in which one
would recover local realism and one of its proponents was Bohm \ \cite%
{Bohmhvt, Omnes, Hughes}. \ Bell developed a clever theoretical tool which
could be used to perform experiments based on it which could decide between
local realism theories and quantum mechanics, the Bell's inequality \cite%
{Bell, Omnes, Hughes, WallsMilburn}. In the 1980s it was realized
experiments by Aspect and collaborators \cite{Aspect, WallsMilburn} based on
the Bell's inequality which although strongly favored quantum mechanics,
there remained some possibility \ for which a local reality view could still
be maintained. We should point out that the main difficulty in the test of
the Bell's inequality is related to the fact that there is no perfect
particle detectors. For a review of Bell's inequalities and some of its
variants see \cite{RevBellsIneq}.

In 1989, Greenberger, Horne and Zeilinger found out an ingenious theoretical
tool to test quantum mechanics confronted with local hidden variable
theories, the Bell theorem without inequalities \cite{GHZ, Mermin,
RevBellsIneq}, which can demonstrate the spookiness of quantum mechanics
even more dramatically than the Bell's analysis. In the GHZ test the
decision between quantum mechanics and local realism theories is a binary
simple one showing the power of the GHZ formalism. \ There are several
proposals of preparation of GHZ states. For instance, proposals involving
cavity QED are presented in \cite{GHZLambdaat, GHZWalther, Gerry, RevHaroche}%
. For a concrete experimental realization of GHZ states see \cite{EXPGHZ,
RevHaroche}. \ In reference \cite{GHZLambdaat} we have proposed a
preparation of atomic GHZ states and a GHZ test similar the one discussed in
this article. However, in reference \cite{GHZLambdaat} we have used
three-level lambda atoms in which the lower states are degenerate states
interacting with a cavity prepared in a coherent state. In the present
article we assume that the atoms are Rydberg atoms of relatively long
radiative lifetimes \cite{Rydat} and that the superconducting\ microwave
cavities \cite{Haroche, Walther} are perfect cavities, that is, we neglect
effects due to decoherence. We are going to consider cavities which are
prepared in a superposition of zero and one Fock states. \ 

\section{ATOMS IN A CASCADE CONFIGURATION\protect\bigskip}

We start showing how to prepare a cavity $C$ in the state 
\begin{equation}
|\psi \rangle _{C}=\frac{(|0\rangle +|1\rangle )}{\sqrt{2}}.  \label{PsiC}
\end{equation}%
In order to prepare this state, we send a two-level atom $A0$, with $%
|f_{0}\rangle $ and $|e_{0}\rangle $ being the lower and upper level
respectively, through a Ramsey cavity $R0$ (see (\ref{URam})) in the lower
state $\mid f_{0}\rangle $ where the atomic states are rotated according to%
\begin{equation}
R_{0}=\frac{1}{\sqrt{2}}\left[ 
\begin{array}{cc}
1 & i \\ 
i & 1%
\end{array}%
\right] ,
\end{equation}%
that is,%
\begin{equation}
\mid f_{0}\rangle \rightarrow \frac{1}{\sqrt{2}}(i\mid e_{0}\rangle +\mid
f_{0}\rangle ),
\end{equation}%
and through $C$, resonant with the cavity. Under the Jaynes-Cummings
dynamics \cite{Orszag} (see\ (\ref{UJC}), for $\Delta =0$). For $gt=\pi /2$
we see that the state $|f_{0}\rangle |0\rangle $ does not evolve, however,
the state $|e_{0}\rangle |0\rangle $ evolves to $-i|f_{0}\rangle |1\rangle $%
. Then, for the cavity initially in the vacuum state $|0\rangle $, we have%
\begin{equation}
\frac{(|f_{0}\rangle +i|e_{0}\rangle )}{\sqrt{2}}|0\rangle \longrightarrow
|f_{0}\rangle \frac{(|0\rangle +|1\rangle )}{\sqrt{2}}=|f_{0}\rangle |\psi
\rangle _{C}.
\end{equation}

Now let us consider a three-level cascade atom \ $Ak$ with $\mid
e_{k}\rangle ,\mid f_{k}\rangle $ and $\mid g_{k}\rangle $ being the upper,
intermediate and lower atomic state respectively (see Fig. 1). We assume
that the transition $\mid f_{k}\rangle \rightleftharpoons \mid e_{k}\rangle $
is far enough from resonance with the cavity central frequency such that
only virtual transitions occur between these states (only these states
interact with field in cavity $C$). In addition we assume that the
transition $\mid e_{k}\rangle \rightleftharpoons \mid g_{k}\rangle $ is
highly detuned from the cavity frequency so that there will be no coupling
with the cavity field. Here we are going to consider the effect of the
atom-field interaction taking into account only levels $\mid f_{k}\rangle $
and $\mid g_{k}\rangle .$ We do not consider level $\mid e_{k}\rangle $
since it will not play any role in our scheme. Therefore, we have
effectively a two-level system involving states $\mid f_{k}\rangle $ and $%
|g_{k}\rangle $. Considering levels $\mid f_{k}\rangle $ and $\mid
g_{k}\rangle ,$ we can write an effective time evolution operator 
\begin{equation}
U_{k}(t)=e^{i\varphi a^{\dagger }a}\mid f_{k}\rangle \langle f_{k}\mid
+|g_{k}\rangle \langle g_{k}\mid ,  \label{UCasc}
\end{equation}%
(see (\ref{Ud})) where the second term above was put by hand just in order
to take into account the effect of level $\mid g_{k}\rangle $ and where $%
\varphi =g^{2}\tau /$ $\Delta $, \ $g$ is the coupling constant, $\Delta
=\omega _{e}-\omega _{f}-\omega $ is the detuning \ where \ $\omega _{e}$
and $\omega _{f}$ \ are the frequencies of the upper and intermediate levels
respectively and $\omega $ is the cavity field frequency and $\tau $ is the
atom-field interaction time. Let us take $\varphi =\pi $. \ Now, let us
assume that we let atom $A1$ to interact with cavity $C$ prepared in the
state (\ref{PsiC}). Let us assume that atom $A1$ is sent to a Ramsey cavity $%
R1$ in the lower state $\mid g_{1}\rangle $ \ where it is prepared cprepared
in a coherent superposition according to the rotation matrix%
\begin{equation}
R_{1}=\frac{1}{\sqrt{2}}\left[ 
\begin{array}{cc}
1 & 1 \\ 
-1 & 1%
\end{array}%
\right] ,
\end{equation}%
and we have 
\begin{equation}
\mid \psi \rangle _{A1}=\frac{1}{\sqrt{2}}(\mid f_{1}\rangle +\mid
g_{1}\rangle ).
\end{equation}%
Taking into account (\ref{UCasc}), after atom $A1$ has passed through the
cavity prepared in state (\ref{PsiC}), we get 
\begin{equation}
\mid \psi \rangle _{A1-C}=\frac{1}{2}[(\mid f_{1}\rangle +\mid g_{1}\rangle
)|0\rangle +(-\mid f_{1}\rangle +\mid g_{1}\rangle )|1\rangle ],
\end{equation}%
Now, if atom $A1$ enters a second Ramsey cavity $R2$ where the atomic states
are rotated according to the rotation matrix%
\begin{equation}
R_{2}=\frac{1}{\sqrt{2}}\left[ 
\begin{array}{cc}
1 & 1 \\ 
-1 & 1%
\end{array}%
\right] ,
\end{equation}%
we have 
\begin{eqnarray}
\frac{1}{\sqrt{2}}( &\mid &f_{1}\rangle +\mid g_{1}\rangle )\rightarrow \mid
f_{1}\rangle ,  \nonumber \\
\frac{1}{\sqrt{2}}(- &\mid &f_{1}\rangle +\mid g_{1}\rangle )\rightarrow
\mid g_{1}\rangle ,
\end{eqnarray}%
and, therefore, 
\begin{equation}
\mid \psi \rangle _{A1-C}=\frac{1}{\sqrt{2}}[\mid f_{1}\rangle |0\rangle
+\mid g_{1}\rangle |1\rangle ],
\end{equation}%
Now, let us prepare a two-level atom $A2$ in the Ramsey cavity $R3$. If atom 
$A2$ is initially in the state $\mid g_{2}\rangle $, according to the
rotation matrix%
\begin{equation}
R_{3}=\frac{1}{\sqrt{2}}\left[ 
\begin{array}{cc}
1 & 1 \\ 
-1 & 1%
\end{array}%
\right] ,
\end{equation}%
we have%
\begin{equation}
\mid \psi \rangle _{A2}=\frac{1}{\sqrt{2}}(\mid f_{2}\rangle +\mid
g_{2}\rangle ),
\end{equation}%
and let us send this atom through cavity $C$, assuming that \ for atom $A2$,
as above for atom $A1$, the transition $\mid f_{2}\rangle \rightleftharpoons
\mid e_{2}\rangle $ is highly detuned from the cavity central frequency.
Taking into account (\ref{UCasc}), after the atom has passed through the
cavity we get 
\begin{equation}
\mid \psi \rangle _{A1-A2-C}=\frac{1}{2}[\mid f_{1}\rangle (\mid
f_{2}\rangle +\mid g_{2}\rangle )|0\rangle +\mid g_{1}\rangle (-\mid
f_{2}\rangle +\mid g_{2}\rangle )|1\rangle ],
\end{equation}%
Then, atom $A2$ enters a Ramsey cavity $R4$ where the atomic states are
rotated according to the rotation matrix%
\begin{equation}
R_{4}=\frac{1}{\sqrt{2}}\left[ 
\begin{array}{cc}
1 & 1 \\ 
-1 & 1%
\end{array}%
\right] ,
\end{equation}%
that is,%
\begin{eqnarray}
\frac{1}{\sqrt{2}}( &\mid &f_{2}\rangle +\mid g_{2}\rangle )\rightarrow \mid
f_{2}\rangle ,  \nonumber \\
\frac{1}{\sqrt{2}}(- &\mid &f_{2}\rangle +\mid g_{2}\rangle )\rightarrow
\mid g_{2}\rangle ,
\end{eqnarray}%
and we get%
\begin{equation}
\mid \psi ,+\rangle _{A1-A2-C}=\frac{1}{\sqrt{2}}[\mid f_{1}\rangle \mid
f_{2}\rangle |0\rangle +\mid g_{1}\rangle \mid g_{2}\rangle |1\rangle ],
\label{GHZCasc+}
\end{equation}%
which is a GHZ state. If we had started with the cavity in the state $%
(|0\rangle -|1\rangle )/\sqrt{2}$ we would get the GHZ state%
\begin{equation}
\mid \psi ,-\rangle _{A1-A2-C}=\frac{1}{\sqrt{2}}[\mid f_{1}\rangle \mid
f_{2}\rangle |0\rangle -\mid g_{1}\rangle \mid g_{2}\rangle |1\rangle ].
\label{GHZCasc-}
\end{equation}

Let us now first discuss a summary of the GHZ test prescription. We will
follow closely the discussion presented in the very clear article by Mermin
\ \cite{Mermin}. First we define the atomic operators%
\begin{eqnarray}
A &=&\sigma _{x}^{1}\sigma _{y}^{2}\sigma _{y}^{3},  \nonumber \\
B &=&\sigma _{y}^{1}\sigma _{x}^{2}\sigma _{y}^{3},  \nonumber \\
C &=&\sigma _{y}^{1}\sigma _{y}^{2}\sigma _{x}^{3},  \nonumber \\
D &=&\sigma _{x}^{1}\sigma _{x}^{2}\sigma _{x}^{3},  \label{ABCD}
\end{eqnarray}%
where%
\begin{eqnarray}
\sigma _{x}^{k} &=&\mid f_{k}\rangle \langle g_{k}\mid +\mid g_{k}\rangle
\langle f_{k}\mid ,  \nonumber \\
\sigma _{y}^{k} &=&-i(\mid f_{k}\rangle \langle g_{k}\mid -\mid g_{k}\rangle
\langle f_{k}\mid ), \\
\sigma _{x}^{3} &=&\mid 0\rangle \langle 1\mid +\mid 1\rangle \langle 0\mid ,
\\
\sigma _{y}^{3} &=&-i(\mid 0\rangle \langle 1\mid -\mid 1\rangle \langle
0\mid ),
\end{eqnarray}%
$(k=1,2$ $)$. It is easy to show that the commutators%
\begin{equation}
\lbrack A,B]=[A,C]=[B,C]=0,
\end{equation}%
and that%
\begin{equation}
A\mid \psi ,\pm \rangle _{A1-A2-C}=B\mid \psi ,\pm \rangle _{A1-A2-C}=C\mid
\psi ,\pm \rangle _{A1-A2-C}=\mp 1\mid \psi ,\pm \rangle _{A1-A2-C},
\end{equation}%
and%
\begin{equation}
D\mid \psi ,\pm \rangle _{A1-A2-C}=\pm 1\mid \psi ,\pm \rangle _{A1-A2-C}.
\label{DPSI}
\end{equation}

If we assume that there are six elements of reality $m_{x}^{k}$ and $%
m_{y}^{k}$ $(k=1,2$ and $3)$ each having value $+1$ or $-1$ waiting to be
revealed according to a local realism theory, then we can write%
\begin{eqnarray}
a_{\pm } &=&m_{x}^{1}m_{y}^{2}m_{y}^{3}=\mp 1,  \nonumber \\
b_{\pm } &=&m_{y}^{1}m_{x}^{2}m_{y}^{3}=\mp 1,  \nonumber \\
c_{\pm } &=&m_{y}^{1}m_{y}^{2}m_{x}^{3}=\mp 1,
\end{eqnarray}%
and we have%
\begin{equation}
d_{\pm }=a_{\pm }b_{\pm }c_{\pm }=m_{x}^{1}m_{x}^{2}m_{x}^{3}=\mp 1,
\end{equation}%
where the upper and lower subindexes refer to the GHZ state (\ref{GHZCasc+})
and (\ref{GHZCasc-}) respectively and we have used $(m_{y}^{k})^{2}=1$. So,
the existence of elements of reality implies that if we measure the value of
the observables $\sigma _{x}^{k}$ $(k=1,2$ and $3)$ (that is the elements of
reality associated with them) in the state $\mid \psi ,\pm \rangle
_{A1-A2-C} $ the product of the three resulting values must be $d_{\pm }=\mp
1$. But according to (\ref{DPSI}) the eigenvalue of the operator $D$ applied
the state $\mid \psi ,\pm \rangle _{A1-A2-C}$ is $\pm 1$. Therefore,
measuring this eigenvalue we can decide between theories based on local
realism and quantum mechanics.

Now, let us see how we proceed to perform the GHZ test (we follow closely
the scheme presented in \cite{GHZLambdaat}). We start letting atoms $Ak$ to
pass through the Ramsey zones $Kk$ \ $(k=1,2$ $)$ where the atomic states
are rotated according to the rotation matrix

\begin{equation}
K_{k}=\frac{1}{\sqrt{2}}\left[ 
\begin{array}{cc}
1 & -1 \\ 
1 & 1%
\end{array}%
\right] ,
\end{equation}%
or%
\begin{equation}
K_{k}=\frac{1}{\sqrt{2}}(\mid f_{k}\rangle \langle f_{k}\mid -\mid
f_{k}\rangle \langle g_{k}\mid +\mid g_{k}\rangle \langle f_{k}\mid +\mid
g_{k}\rangle \langle g_{k}\mid ).  \label{Kk}
\end{equation}%
These rotation matrixes are the key ingredient which allow us to perform the
GHZ test. The method of the test we are going to describe is based on a
gradual unraveling of the GHZ state being considered.

The eigenvalues of the operators $\sigma _{x}^{k}$ are%
\begin{equation}
|\psi _{x}^{k},\pm \rangle =\frac{1}{\sqrt{2}}(\mid f_{k}\rangle \pm \mid
g_{k}\rangle ).  \label{PSIx}
\end{equation}%
for $k=1,2.$ Let us take the state (\ref{GHZCasc+}). Writing the states $%
\mid f_{1}\rangle $ and $\mid g_{1}\rangle $ in terms of the states (\ref%
{PSIx}) for $k=1$ and substituting in the GHZ state (\ref{GHZCasc+}) we get

\begin{equation}
\mid \psi ,+\rangle _{A1-A2-C}=\frac{1}{2}[|\psi _{x}^{1},+\rangle (\mid
f_{2}\rangle \mid 0\rangle +\mid g_{2}\rangle \mid 1\rangle )+|\psi
_{x}^{1},-\rangle (\mid f_{2}\rangle \mid 0\rangle -\mid g_{2}\rangle \mid
1\rangle )].  \label{PSIX1}
\end{equation}%
Applying (\ref{Kk}) in the state (\ref{GHZCasc+}) for $k=1$, we have%
\begin{equation}
K_{1}\mid \psi ,+\rangle _{A1-A2-C}=\frac{1}{2}[|f_{1}\rangle (\mid
f_{2}\rangle \mid 0\rangle -\mid g_{2}\rangle \mid 1\rangle )+|g_{1}\rangle
(\mid f_{2}\rangle \mid 0\rangle +\mid g_{2}\rangle \mid 1\rangle )].
\label{K1PSI123}
\end{equation}%
Now, we compare (\ref{K1PSI123}) and (\ref{PSIX1}). We see that the rotation
by $K_{1}$ followed by the detection of $|g_{1}\rangle $ corresponds to the
detection of the the state $|\psi _{x}^{1},+\rangle $ whose eigenvalue of $%
\sigma _{x}^{1}$ is $+1$. After we detect $|g_{1}\rangle $, we get%
\begin{equation}
\mid \psi \rangle _{A2-C}=\frac{1}{\sqrt{2}}(\mid f_{2}\rangle \mid 0\rangle
+\mid g_{2}\rangle \mid 1\rangle ).  \label{PSI23}
\end{equation}%
As before, we rewrite the states $\mid f_{2}\rangle $ and $\mid g_{2}\rangle 
$ in terms of these states (\ref{PSIx}) for $k=2$ and substitute in the
above state and we have 
\begin{equation}
\mid \psi \rangle _{A2-C}=\frac{1}{2}[|\psi _{x}^{2},+\rangle (\mid 0\rangle
+\mid 1\rangle )+|\psi _{x}^{2},-\rangle (\mid 0\rangle -\mid 1\rangle )].
\label{PSIX2}
\end{equation}%
If we apply (\ref{Kk}) for $k=2$ to the state (\ref{PSI23}) we get%
\begin{equation}
K_{2}\mid \psi \rangle _{A2-C}=\frac{1}{2}[|f_{2}\rangle (\mid 0\rangle
-\mid 1\rangle )+|g_{2}\rangle (\mid 0\rangle +\mid 1\rangle )].
\label{K2PSI23}
\end{equation}%
Again, if we compare (\ref{K2PSI23}) with (\ref{PSIX2}), we see that the
rotation by $K_{2}$ followed by the detection of $|g_{2}\rangle $
corresponds to the detection of the the state $|\psi _{x}^{2},+\rangle $
whose eigenvalue of $\sigma _{x}^{2}$ is $+1$. After we detect $%
|g_{2}\rangle $, we get%
\begin{equation}
|\psi _{x}^{3},+\rangle =\frac{1}{\sqrt{2}}(\mid 0\rangle +\mid 1\rangle ),
\label{PSIA3}
\end{equation}%
which is the eigenvector of $\sigma _{x}^{3}$ with eigenvalue $+1$. We now
send a two-level atom $A3,$ resonant with the cavity, with $|g_{3}\rangle $
and $|f_{3}\rangle $ being the lower and upper level respectively, through $%
C $. If $A3$ is sent in the lower state $|g_{3}\rangle $, under the
Jaynes-Cummings dynamics, for $gt=\pi /2,$ we know that the state $%
|g_{3}\rangle |0\rangle $ does not evolve, however, the state $|g_{3}\rangle
|1\rangle $ evolves to $-i|f_{3}\rangle |0\rangle $. Then we get%
\begin{equation}
|g_{3}\rangle |\psi _{x}^{3},+\rangle \longrightarrow \frac{(|g_{3}\rangle
-i|f_{3}\rangle )}{\sqrt{2}}|0\rangle
\end{equation}%
Now we let atom $A3$ to enter a Ramsey cavity $K3$ where the atomic states
are rotated according to the rotation matrix%
\begin{equation}
K_{3}=\frac{1}{\sqrt{2}}\left[ 
\begin{array}{cc}
1 & i \\ 
i & 1%
\end{array}%
\right] ,  \label{K3}
\end{equation}%
that is,%
\begin{equation}
\frac{1}{\sqrt{2}}(-i\mid f_{3}\rangle +\mid g_{3}\rangle )\rightarrow \mid
g_{3}\rangle ,
\end{equation}%
and we see that the rotation by $K_{3}$ followed by the detection of $%
|g_{3}\rangle $ corresponds to the detection of the the state $|\psi
_{x}^{2},+\rangle $ whose eigenvalue of $\sigma _{x}^{3}$ is $+1$.

In the case of 
\begin{equation}
|\psi _{x}^{3},-\rangle =\frac{1}{\sqrt{2}}(|0\rangle -|1\rangle )
\end{equation}%
which is the eigenvector of $\sigma _{x}^{3}$ with eigenvalue $-1$, we would
get%
\begin{equation}
|g_{3}\rangle |\psi _{x}^{3},-\rangle \longrightarrow \frac{(|g_{3}\rangle
+i|f_{3}\rangle )}{\sqrt{2}}|0\rangle
\end{equation}%
and after the rotation (\ref{K3}) we would have%
\begin{equation}
\frac{1}{\sqrt{2}}(i\mid f_{3}\rangle +\mid g_{3}\rangle )\rightarrow i\mid
f_{3}\rangle .
\end{equation}%
and we see that the rotation by $K_{3}$ followed by the detection of $%
|f_{3}\rangle $ corresponds to the detection of the the state $|\psi
_{x}^{2},-\rangle $ whose eigenvalue of $\sigma _{x}^{3}$ is $-1$.

We can repeat the above procedure and see that we have only four
possibilities which are presented schematically below, where on the left, we
present the possible sequences of atomic state rotations through $K_{k}$ and
detections of $\mid f_{j}\rangle $ or $\mid g_{j}\rangle $ $(j=1,2)$ and $%
\mid g_{3}\rangle $ or $\mid f_{3}\rangle $ and on the right, we present the
sequences of the corresponding states $|\psi _{x}^{k},\pm \rangle $ where $%
k=1,2$ and $3$ which corresponds to the measurement of the eigenvalue of the
operator $D$ given by (\ref{ABCD}), 
\begin{eqnarray}
(K_{1}, &\mid &g_{1}\rangle )(K_{2},\mid g_{2}\rangle )(K_{3},\mid
g_{3}\rangle )\longleftrightarrow |\psi _{x}^{1},+\rangle |\psi
_{x}^{2},+\rangle |\psi _{x}^{3},+\rangle ,  \nonumber \\
(K_{1}, &\mid &g_{1}\rangle )(K_{2},\mid f_{2}\rangle )(K_{3},\mid
f_{3}\rangle )\longleftrightarrow |\psi _{x}^{1},+\rangle |\psi
_{x}^{2},-\rangle |\psi _{x}^{3},-\rangle ,  \nonumber \\
(K_{1}, &\mid &f_{1}\rangle )(K_{2},\mid f_{2}\rangle )(K_{3},\mid
g_{3}\rangle )\longleftrightarrow |\psi _{x}^{1},-\rangle |\psi
_{x}^{2},-\rangle |\psi _{x}^{3},+\rangle ,  \nonumber \\
(K_{1}, &\mid &f_{1}\rangle )(K_{2},\mid g_{2}\rangle )(K_{3},\mid
f_{3}\rangle )\longleftrightarrow |\psi _{x}^{1},-\rangle |\psi
_{x}^{2},+\rangle |\psi _{x}^{3},-\rangle .  \label{GHZT+}
\end{eqnarray}%
Notice that all these results give us the eigenvalue $+1$ of the operator $D$
(see (\ref{DPSI})). Therefore, in one run, if we succeed to perform the
above rotations of the atomic states according to (\ref{Kk}), each one
followed by the detection of the respective state $\mid f_{j}\rangle $ or $%
\mid g_{j}\rangle $ $(j=1,2)$ and $\mid g_{3}\rangle $ or $\mid f_{3}\rangle 
$, we get the result of the experiment in favor of quantum mechanics.

If we perform the test for the state (\ref{GHZCasc-}), following the above
procedure, it is easy to see that we have the four possible outcomes,%
\begin{eqnarray}
(K_{1}, &\mid &g_{1}\rangle )(K_{2},\mid g_{2}\rangle )(K_{3},\mid
f_{3}\rangle )\longleftrightarrow |\psi _{x}^{1},+\rangle |\psi
_{x}^{2},+\rangle |\psi _{x}^{3},-\rangle ,  \nonumber \\
(K_{1}, &\mid &g_{1}\rangle )(K_{2},\mid f_{2}\rangle )(K_{3},\mid
g_{3}\rangle )\longleftrightarrow |\psi _{x}^{1},+\rangle |\psi
_{x}^{2},-\rangle |\psi _{x}^{3},+\rangle ,  \nonumber \\
(K_{1}, &\mid &f_{1}\rangle )(K_{2},\mid f_{2}\rangle )(K_{3},\mid
f_{3}\rangle )\longleftrightarrow |\psi _{x}^{1},-\rangle |\psi
_{x}^{2},-\rangle |\psi _{x}^{3},-\rangle ,  \nonumber \\
(K_{1}, &\mid &f_{1}\rangle )(K_{2},\mid g_{2}\rangle )(K_{3},\mid
g_{3}\rangle )\longleftrightarrow |\psi _{x}^{1},-\rangle |\psi
_{x}^{2},+\rangle |\psi _{x}^{3},+\rangle .  \label{GHZT-}
\end{eqnarray}%
Notice that all these results give us the eigenvalue $-1$ of the operator $D$
(see (\ref{DPSI})), as it should be. See Fig. 2 where we present a scheme of
the GHZ experiment.

\section{ATOMS IN A LAMBDA CONFIGURATION\protect\bigskip}

Consider a three-level lambda atom (see Fig. 3) interacting with the
electromagnetic field inside a cavity $C$. The states of \ the atom, $%
|a\rangle ,$ $|b\rangle $ and $|c\rangle $ are so that the $|a\rangle
\rightleftharpoons |c\rangle $ and $|a\rangle \rightleftharpoons |b\rangle $
transitions are in the far off resonance interaction limit. The time
evolution operator for the atom-field interaction $U(t)$ is given by \cite%
{Knight} 
\begin{equation}
U(\tau )=-e^{i\varphi a^{\dagger }a}|a\rangle \langle a|+\frac{1}{2}%
(e^{i\varphi a^{\dagger }a}+1)|b\rangle \langle b|+\frac{1}{2}(e^{i\varphi
a^{\dagger }a}-1)|b\rangle \langle c|\ +\frac{1}{2}(e^{i\varphi a^{\dagger
}a}-1)|c\rangle \langle b|+\frac{1}{2}(e^{i\varphi a^{\dagger
}a}+1)|c\rangle \langle c|,  \label{U1lambda}
\end{equation}%
where $a$ $(a^{\dagger })$ is the annihilation (creation) operator for the
field in cavity $C$, $\varphi =2g^{2}\tau /$ $\Delta $, \ $g$ is the
coupling constant, $\Delta =\omega _{a}-\omega _{b}-\omega =\omega
_{a}-\omega _{c}-\omega $ is the detuning where \ $\omega _{a}$, $\omega
_{b} $ and $\omega _{c}$\ are the frequency of the upper level and \ of the
two degenerate lower levels respectively and $\omega $ is the cavity field
frequency and $\tau $ is the atom-field interaction time. For $\varphi =\pi $%
, we get 
\begin{equation}
U(\tau )=-\exp \left( i\pi a^{\dagger }a\right) |a\rangle \langle a|+\Pi
_{+}|b\rangle \langle b|+\Pi _{-}|b\rangle \langle c|\ +\Pi _{-}|c\rangle
\langle b|+\Pi _{+}|c\rangle \langle c|,  \label{UlambdaPi}
\end{equation}%
where 
\begin{eqnarray}
\Pi _{+} &=&\frac{1}{2}(e^{i\pi a^{\dagger }a}+1),  \nonumber \\
\Pi _{-} &=&\frac{1}{2}(e^{i\pi a^{\dagger }a}-1).  \label{pi+-}
\end{eqnarray}

Let us first show how we can get a \ GHZ state making use of three-level
lambda atoms interacting with a cavity field prepared in state (\ref{PsiC}).
Consider the atom $A1$ in the state $|\psi \rangle _{A1}=|b_{1}\rangle $ and
a cavity $C$ prepared in the state (\ref{PsiC}). We now let atom $A1$ fly
through the cavity $C$. Taking into account (\ref{UlambdaPi}) the state \ of
the system $A1-C$ evolves to 
\begin{equation}
|\psi \rangle _{A1-C}==\frac{1}{\sqrt{2}}(|b_{1}\rangle |0\rangle
-|c_{1}\rangle |1\rangle ).
\end{equation}%
Consider now another three-level lambda atom $A2$ prepared initially in the
state $|b_{2}\rangle $, which are going to pass through the cavity. After
this second atom has passed through the cavity, the system evolves to 
\begin{equation}
|\psi ,+\rangle _{A1-A2-C}=\frac{1}{\sqrt{2}}(|b_{1}\rangle |b_{2}\rangle
|0\rangle +|c_{1}\rangle |c_{2}\rangle |1\rangle ).  \label{GHZLamb+}
\end{equation}%
which is a a GHZ state. If we had started with the cavity in the state $%
(|0\rangle -|1\rangle )/\sqrt{2}$ we would get the GHZ state%
\begin{equation}
\mid \psi ,-\rangle _{A1-A2-C}=\frac{1}{\sqrt{2}}[\mid b_{1}\rangle \mid
b_{2}\rangle |0\rangle -\mid c_{1}\rangle \mid c_{2}\rangle |1\rangle ].
\label{GHZLamb-}
\end{equation}

In the case of (\ref{GHZLamb+}), following the prescription of the previous
section, we can see that in the GHZ test we have only four possibilities
which are presented schematically below, where on the left, we present the
possible sequences of atomic state rotations through $K_{k}$ and detections
of $\mid b_{j}\rangle $ or $\mid c_{j}\rangle $ $(j=1,2)$ and $\mid
g_{3}\rangle $ or $\mid f_{3}\rangle $ and on the right, we present the
sequences of the corresponding states $|\psi _{x}^{k},\pm \rangle $ where $%
k=1,2$ and $3$ which corresponds to the measurement of the eigenvalue of the
operator $D$ given by (\ref{ABCD}), 
\begin{eqnarray}
(K_{1}, &\mid &c_{1}\rangle )(K_{2},\mid c_{2}\rangle )(K_{3},\mid
g_{3}\rangle )\longleftrightarrow |\psi _{x}^{1},+\rangle |\psi
_{x}^{2},+\rangle |\psi _{x}^{3},+\rangle ,  \nonumber \\
(K_{1}, &\mid &c_{1}\rangle )(K_{2},\mid b_{2}\rangle )(K_{3},\mid
f_{3}\rangle )\longleftrightarrow |\psi _{x}^{1},+\rangle |\psi
_{x}^{2},-\rangle |\psi _{x}^{3},-\rangle ,  \nonumber \\
(K_{1}, &\mid &b_{1}\rangle )(K_{2},\mid b_{2}\rangle )(K_{3},\mid
g_{3}\rangle )\longleftrightarrow |\psi _{x}^{1},-\rangle |\psi
_{x}^{2},-\rangle |\psi _{x}^{3},+\rangle ,  \nonumber \\
(K_{1}, &\mid &b_{1}\rangle )(K_{2},\mid c_{2}\rangle )(K_{3},\mid
f_{3}\rangle )\longleftrightarrow |\psi _{x}^{1},-\rangle |\psi
_{x}^{2},+\rangle |\psi _{x}^{3},-\rangle .
\end{eqnarray}%
where now,%
\begin{equation}
|\psi _{x}^{k},\pm \rangle =\frac{1}{\sqrt{2}}(\mid b_{k}\rangle \pm \mid
c_{k}\rangle ),
\end{equation}%
for $k=1,2$. Notice that all these results give us the eigenvalue $+1$ of
the operator $D$ (see (\ref{DPSI})). Therefore, in one run, if we succeed to
perform the above rotations of the atomic states according to (\ref{Kk}),
each one followed by the detection of the respective state $\mid
b_{j}\rangle $ or $\mid c_{j}\rangle $ $(j=1,2)$ and $\mid g_{3}\rangle $ or 
$\mid f_{3}\rangle $, we get the result of the experiment in favor of
quantum mechanics.

If we perform the test for the state (\ref{GHZLamb-}), following the above
procedure, it is easy to see that we have the four possible outcomes,%
\begin{eqnarray}
(K_{1}, &\mid &c_{1}\rangle )(K_{2},\mid c_{2}\rangle )(K_{3},\mid
f_{3}\rangle )\longleftrightarrow |\psi _{x}^{1},+\rangle |\psi
_{x}^{2},+\rangle |\psi _{x}^{3},-\rangle ,  \nonumber \\
(K_{1}, &\mid &c_{1}\rangle )(K_{2},\mid b_{2}\rangle )(K_{3},\mid
g_{3}\rangle )\longleftrightarrow |\psi _{x}^{1},+\rangle |\psi
_{x}^{2},-\rangle |\psi _{x}^{3},+\rangle ,  \nonumber \\
(K_{1}, &\mid &b_{1}\rangle )(K_{2},\mid b_{2}\rangle )(K_{3},\mid
f_{3}\rangle )\longleftrightarrow |\psi _{x}^{1},-\rangle |\psi
_{x}^{2},-\rangle |\psi _{x}^{3},-\rangle ,  \nonumber \\
(K_{1}, &\mid &b_{1}\rangle )(K_{2},\mid c_{2}\rangle )(K_{3},\mid
g_{3}\rangle )\longleftrightarrow |\psi _{x}^{1},-\rangle |\psi
_{x}^{2},+\rangle |\psi _{x}^{3},+\rangle .
\end{eqnarray}%
Notice that all these results give us the eigenvalue $-1$ of the operator $D$
(see (\ref{DPSI})), as it should be. See Fig. 4 where we present a scheme of
the GHZ experiment.

\section{CONCLUSION\protect\bigskip}

We have presented two schemes of preparation of GHZ states and realization
of the GHZ test via cavity QED. In these schemes the cavity is prepared in a
superposition of zero and one Fock states which are not affected
dramatically by decoherence and which are relatively easy to handle. In the
first scheme we make use of atoms in a cascade configuration and in the
second scheme we make use of atoms in a three-level lambda configuration.
The advantage of using a cascade atomic configuration is that the detection
process is easier, however, we have to perform many atomic state rotations.
The advantage of using the lambda atomic configuration is that we do not
have to use many atomic state rotations, however, the detection of the
atomic states are more complicated.

\bigskip

\appendix

\section{Time evolution operator}

\bigskip

\subsection{Two-level atoms\protect\bigskip}

Let us consider a two-level atom interacting with a cavity field, where $%
|e\rangle $ and $|f\rangle $ are the upper and lower states respectively,
with $\omega _{e}$ and $\omega _{f}$ being the two atomic frequencies
associated to these two states and $\omega $ the cavity field frequency (see
Fig. 1). The Jaynes-Cummings Hamiltonian, under the rotating-wave
approximation, is given by 
\begin{equation}
H=\hbar a^{\dag }a+\hbar \omega _{e}|e\rangle \langle e|++\hbar \omega
_{f}|f\rangle \langle f|+\hbar g[a|e\rangle \langle f|+a^{\dag }|f\rangle
\langle e|),  \label{JCH}
\end{equation}%
where $a^{\dag }$ and $a$ are the creation and annihilation operators
respectively for the cavity field, $g$ is the coupling constant and we write%
\begin{equation}
H=H_{0}+H_{I},
\end{equation}%
where we have settled%
\begin{eqnarray}
H_{0} &=&\hbar a^{\dag }a+\hbar \omega _{e}|e\rangle \langle e|++\hbar
\omega _{f}|f\rangle \langle f|,  \nonumber \\
H_{I} &=&\hbar g(a|e\rangle \langle f|+a^{\dag }|f\rangle \langle e|).
\end{eqnarray}%
Lets define the interaction picture%
\begin{equation}
|\psi _{I}\rangle =e^{i\frac{H_{0}}{\hbar }t}|\psi _{S}\rangle ,
\end{equation}%
where%
\begin{equation}
i\hbar \frac{d}{dt}|\psi _{S}\rangle =H|\psi _{S}\rangle .
\end{equation}%
Then, we get%
\begin{equation}
i\hbar \frac{d}{dt}|\psi _{I}\rangle =V_{I}|\psi _{I}\rangle ,
\end{equation}%
where%
\begin{equation}
V_{I}=e^{i\frac{H_{0}}{\hbar }t}H_{I}e^{-i\frac{H_{0}}{\hbar }t}=\hbar \left[
\begin{array}{cc}
0 & ge^{i\Delta t}a \\ 
ge^{-i\Delta t}a^{\dagger } & 0%
\end{array}%
\right] 
\end{equation}%
and%
\begin{equation}
\Delta =(\omega _{e}-\omega _{f})-\omega .
\end{equation}%
Considering%
\begin{equation}
|\psi _{I}(t)\rangle =U_{I}(t)|\psi _{I}(0)\rangle =U_{I}(t)|\psi
_{S}(0)\rangle ,
\end{equation}%
we have to solve the Schr\"{o}dinger's equation for the time evolution
operator%
\begin{equation}
i\hbar \frac{dU_{I}}{dt}=V_{I}U_{I},  \label{AP2}
\end{equation}%
where%
\begin{equation}
U_{I}(t)=\left[ 
\begin{array}{cc}
u_{ee}(t) & u_{ef}(t) \\ 
u_{fe}(t) & u_{ff}(t)%
\end{array}%
\right] 
\end{equation}%
and%
\begin{equation}
U_{I}(0)=\left[ 
\begin{array}{cc}
1 & 0 \\ 
0 & 1%
\end{array}%
\right] .
\end{equation}%
That is,%
\begin{eqnarray}
i\frac{d}{dt}u_{ee}(t) &=&ge^{i\Delta t}au_{ef}(t),  \nonumber \\
i\frac{d}{dt}u_{ef}(t) &=&ge^{i\Delta t}au_{ff}(t),  \nonumber \\
i\frac{d}{dt}u_{fe}(t) &=&ge^{-i\Delta t}a^{\dagger }u_{ee}(t),  \nonumber \\
i\frac{d}{dt}u_{ff}(t) &=&ge^{-i\Delta t}a^{\dagger }u_{ef}(t),  \label{EDI4}
\end{eqnarray}%
which can be solved easily using, for instance, Laplace transformation, and
we get%
\begin{equation}
U_{I}(t)=\left[ 
\begin{array}{cc}
e^{i\frac{\Delta }{2}t}(\cos \mu t-i\frac{\Delta }{2\mu }\sin \mu t) & 
-ige^{i\frac{\Delta }{2}t}\frac{1}{\mu }(\sin \mu t)a \\ 
-iga^{\dag }e^{-i\frac{\Delta }{2}t}\frac{1}{\mu }(\sin \mu t) & e^{-i\frac{%
\Delta }{2}t}(\cos \nu t+i\frac{\Delta }{2\nu }\sin \nu t)%
\end{array}%
\right] ,  \label{UJC}
\end{equation}%
where we have defined%
\begin{eqnarray}
\mu  &=&\sqrt{g^{2}aa^{\dag }+\frac{\Delta ^{2}}{4}},  \nonumber \\
\nu  &=&\sqrt{g^{2}a^{\dag }a+\frac{\Delta ^{2}}{4}}.  \label{AP4}
\end{eqnarray}

In the large detuning limit ($\Delta \gg g$) we have%
\begin{eqnarray}
\mu &=&\sqrt{g^{2}aa^{\dag }+\frac{\Delta ^{2}}{4}}\cong \frac{\Delta }{2}+%
\frac{g^{2}aa^{\dag }}{\Delta },  \nonumber \\
\nu &=&\sqrt{g^{2}a^{\dag }a+\frac{\Delta ^{2}}{4}}.\cong \frac{\Delta }{2}+%
\frac{g^{2}a^{\dag }a}{\Delta }.
\end{eqnarray}%
and we get easily 
\begin{equation}
U_{d}(t)=e^{-i\varphi (a^{\dagger }a+1)}\mid e\rangle \langle e\mid
+e^{i\varphi a^{\dagger }a}\mid f\rangle \langle f\mid ,  \label{Ud}
\end{equation}%
where $\varphi =$ $g^{2}t/\Delta .$ If we neglect spontaneous emission, the
above time evolution operator can also be obtained from the effective
Hamiltonian%
\begin{equation}
H_{d}=\hbar \frac{g^{2}}{\Delta }a^{\dagger }a(\mid e\rangle \langle e\mid
-\mid f\rangle \langle f\mid ).  \label{Hd}
\end{equation}%
The subindexes in (\ref{Ud}) and (\ref{Hd}) are related to the atom-field
interaction described by them, that is, a dispersive interaction.

In the case we have a resonant interaction of an atom with cavity field ($%
\Delta =0$ in (\ref{UJC})), if the field is a very intense field we can
treat it classically. That is, in the time evolution operator (\ref{UJC}) \
we set $\Delta =0,$ and we substitute the creation and annihilation field
operators according to $a\rightarrow \eta e^{i\theta }$ and $a^{\dag
}\rightarrow \eta e^{-i\theta }$ where $\eta $ and $\theta $ are c-numbers.
Then, we have a semiclassical approach of the atom-field interaction in
which the field is treated classically and the atoms according to quantum
mechanics. In this case (\ref{UJC}) becomes%
\begin{equation}
U_{I,SC}(t)=\left[ 
\begin{array}{cc}
\cos (g\eta t) & -ie^{i\theta }\sin (g\eta t) \\ 
-ie^{-i\theta }\sin (g\eta t) & \cos (g\eta t)%
\end{array}%
\right] .  \label{URam}
\end{equation}%
Now we take $\theta =\pi /2$. If we choose $g\eta t=\pi /4$ we have the
rotation matrix%
\begin{equation}
R_{\frac{\pi }{2},\frac{\pi }{4}}=\frac{1}{\sqrt{2}}\left[ 
\begin{array}{cc}
1 & 1 \\ 
-1 & 1%
\end{array}%
\right] ,
\end{equation}%
and for $g\eta t=-\pi /4$ we have%
\begin{equation}
R_{\frac{\pi }{2},-\frac{\pi }{4}}=\frac{1}{\sqrt{2}}\left[ 
\begin{array}{cc}
1 & -1 \\ 
1 & 1%
\end{array}%
\right] .
\end{equation}%
For $g\eta t=\pi /2$ we get 
\begin{equation}
R_{\frac{\pi }{2},\frac{\pi }{2}}=\left[ 
\begin{array}{cc}
0 & 1 \\ 
-1 & 0%
\end{array}%
\right] ,
\end{equation}%
and for $g\eta t=-\pi /2$ 
\begin{equation}
R_{\frac{\pi }{2},-\frac{\pi }{2}}=\left[ 
\begin{array}{cc}
0 & -1 \\ 
1 & 0%
\end{array}%
\right] .
\end{equation}%
Now if we take $\theta =\pi $ and $g\eta t=\pi /4$ we have the rotation
matrix%
\begin{equation}
R_{\pi ,\frac{\pi }{4}}=\frac{1}{\sqrt{2}}\left[ 
\begin{array}{cc}
1 & i \\ 
i & 1%
\end{array}%
\right] ,
\end{equation}%
Choosing the proper values of $\theta $, $g$, $\eta $ and $t$ we can get the
rotation matrix we need to perform the rotation of the atomic states we
desire in a Ramsey cavity. In all the sections of this article when we
mention a rotation of atomic states in a Ramsey cavity we are taking into
account (\ref{URam}).

Just as a remark, consider (\ref{AP5}) and assume we have an intense field,
that is, we can use a semiclassical approach. In this case we set $%
a\rightarrow \eta e^{i\theta }$ and $a^{\dag }\rightarrow \eta e^{-i\theta }$
where $\eta $ and $\theta $ are c-numbers and $a^{\dag }a\rightarrow \eta
^{2}$. Defining $\varphi a^{\dagger }a\rightarrow \varphi \eta ^{2}=\beta /2$%
, (\ref{AP5}) reads%
\begin{equation}
U_{d,SC}=\left[ 
\begin{array}{cc}
e^{-i\beta /2} & 0 \\ 
0 & e^{i\beta /2}%
\end{array}%
\right] ,  \label{Ud,SC}
\end{equation}%
and for $\beta =\pi $ we have%
\begin{equation}
U_{\pi }=i\left[ 
\begin{array}{cc}
-1 & 0 \\ 
0 & 1%
\end{array}%
\right] .
\end{equation}%
Any arbitrary $2\times 2$ unitary matrix \ $M$\ may be decomposed as \cite%
{Nielsen} 
\begin{equation}
M=e^{i\alpha }\left[ 
\begin{array}{cc}
e^{-i\beta /2} & 0 \\ 
0 & e^{i\beta /2}%
\end{array}%
\right] \left[ 
\begin{array}{cc}
\cos \frac{\gamma }{2} & -\sin \frac{\gamma }{2} \\ 
\sin \frac{\gamma }{2} & \cos \frac{\gamma }{2}%
\end{array}%
\right] \left[ 
\begin{array}{cc}
e^{-i\delta /2} & 0 \\ 
0 & e^{i\delta /2}%
\end{array}%
\right] ,
\end{equation}%
where $\alpha ,\beta ,\gamma $ and $\delta $ are real parameters. Therefore,
we can use (\ref{URam}) and (\ref{Ud,SC}) to get a rotation matrix we need.

\subsection{Three-level lambda atoms\protect\bigskip }

We start with the Hamiltonian of a degenerate three-level lambda atom (see
Fig. 3) interacting with a field cavity mode

\begin{equation}
H=\hbar \omega a^{\dagger }a+\hbar \omega _{a}|a\rangle \langle a|\ +\hbar
\omega _{b}|b\rangle \langle b|+\hbar \omega _{c}|c\rangle \langle c|\
+\hbar a(g_{1}|a\rangle \langle b|+g_{2}|a\rangle \langle c|)+\hbar
a^{\dagger }(g_{1}^{\ast }|b\rangle \langle a|+g_{2}^{\ast }|c\rangle
\langle a|),
\end{equation}%
where $|a\rangle $, $|b\rangle $ and $|c\rangle $ are the upper and the two
degenerated lower atomic levels respectively, $a$ ($a^{\dagger }$) is the
annihilation (creation) field operator and $g_{1}$ and $g_{2}$ are the
coupling constants corresponding to the transitions $|a\rangle
\rightleftharpoons |c\rangle $ and $|a\rangle \rightleftharpoons |b\rangle $%
, respectively. In the interaction picture, 
\begin{equation}
V=\hbar \left[ e^{i\Delta _{1}t}g_{1}a\mid a\rangle \langle b\mid
+e^{-i\Delta _{1}t}g_{1}^{\ast }a^{\dagger }\mid b\rangle \langle a\mid %
\right] +\hbar \left[ e^{i\Delta _{2}t}g_{2}a\mid a\rangle \langle c\mid
+e^{-i\Delta _{2}t}g_{2}^{\ast }a^{\dagger }\mid c\rangle \langle a\mid %
\right] ,
\end{equation}%
where 
\begin{eqnarray}
\Delta _{1} &=&\omega _{a}-\omega _{b}-\omega \ \   \nonumber \\
\ \Delta _{2} &=&\omega _{a}-\omega _{c}-\omega .
\end{eqnarray}%
The time evolution operator 
\begin{equation}
U(t)=\left[ 
\begin{array}{lll}
u_{aa} & u_{ab} & u_{ac} \\ 
u_{ba} & u_{bb} & u_{bc} \\ 
u_{aa} & u_{cb} & u_{cc}%
\end{array}%
\right] ,
\end{equation}%
whose initial condition is given by 
\begin{equation}
U(0)=\left[ 
\begin{array}{lll}
1 & 0 & 0 \\ 
0 & 1 & 0 \\ 
0 & 0 & 1%
\end{array}%
\right] ,
\end{equation}%
should satisfy the Schr\"{o}dinger equation of motion 
\begin{equation}
i\hbar \frac{dU}{dt}=VU=\left[ 
\begin{array}{lll}
\zeta _{1}u_{ba}+\zeta _{2}u_{ca} & \zeta _{1}u_{bb}+\zeta _{2}u_{cb} & 
\zeta _{1}u_{bc}+\zeta _{2}u_{cc} \\ 
\zeta _{1}^{\dagger }u_{aa} & \zeta _{1}^{\dagger }u_{ab} & \zeta
_{1}^{\dagger }u_{ac} \\ 
\zeta _{2}^{\dagger }u_{aa} & \zeta _{2}^{\dagger }u_{ab} & \zeta
_{2}^{\dagger }u_{ac}%
\end{array}%
\right] ,
\end{equation}%
where 
\begin{eqnarray}
\zeta _{1} &=&\hbar e^{i\Delta _{1}t}g_{1}a,  \nonumber \\
\zeta _{2} &=&\hbar e^{i\Delta _{2}t}g_{2}a.
\end{eqnarray}%
Observe that this equation may be grouped in three sets of couple
differential equations. One for $u_{aa}$, $u_{ba}$ and $u_{ca}$, i.e., 
\begin{eqnarray}
i\hbar \frac{du_{aa}}{dt} &=&\zeta _{1}u_{ba}+\zeta _{2}u_{ca},\hspace{8pt} 
\nonumber \\
i\hbar \frac{du_{ba}}{dt} &=&\zeta _{1}^{\dagger }u_{aa},  \nonumber \\
i\hbar \frac{du_{ca}}{dt} &=&\zeta _{2}^{\dagger }u_{aa},
\end{eqnarray}%
another involving only $u_{ab}$, $u_{bb}$ and $u_{cb}$, 
\begin{eqnarray}
i\hbar \frac{du_{ab}}{dt} &=&\zeta _{1}u_{bb}+\zeta _{2}u_{cb},\hspace{8pt} 
\nonumber \\
i\hbar \frac{du_{bb}}{dt} &=&\zeta _{1}^{\dagger }u_{ab},  \nonumber \\
i\hbar \frac{du_{cb}}{dt} &=&\zeta _{2}^{\dagger }u_{ab},
\end{eqnarray}%
and, finally, one involving $u_{ac}$, $u_{bc}$ and $u_{cc}$, 
\begin{eqnarray}
i\hbar \frac{du_{ac}}{dt} &=&\zeta _{1}u_{bc}+\zeta _{2}u_{cc},\hspace{8pt} 
\nonumber \\
i\hbar \frac{du_{bc}}{dt} &=&\zeta _{1}^{\dagger }u_{ac},  \nonumber \\
i\hbar \frac{du_{cc}}{dt} &=&\zeta _{2}^{\dagger }u_{ac}.
\end{eqnarray}

Let us take $\Delta _{1}=\Delta _{2}=\Delta =-i\eta $%
\begin{eqnarray}
\zeta _{1} &=&\hbar e^{i\Delta _{1}t}g_{1}a=\hbar \alpha _{1}e^{\eta t}, 
\nonumber \\
\zeta _{2} &=&\hbar e^{i\Delta _{2}t}g_{2}a=\hbar \alpha _{2}e^{\eta t}.
\end{eqnarray}%
Notice that all the 3 system of differential equations above are of the form 
\begin{eqnarray}
i\frac{dx}{dt} &=&\alpha _{1}e^{\eta t}y+\alpha _{2}e^{\eta t}z,  \nonumber
\\
ie^{\eta t}\frac{dy}{dt} &=&\alpha _{1}^{\dagger }x,  \nonumber \\
ie^{\eta t}\frac{dz}{dt} &=&\alpha _{2}^{\dagger }x.
\end{eqnarray}%
We can solve the systems of differential equations using, for instance,
Laplace transformation and we have for the degenerate case 
\begin{eqnarray}
u_{aa}(t) &=&\frac{e^{i\frac{\Delta t}{2}}}{\sqrt{\mu }}\left[ \sqrt{\mu }%
\cos \sqrt{\mu }t-i\frac{\Delta }{2}\sin \sqrt{\mu }t\right] ,  \nonumber \\
u_{ab}(t) &=&-i\frac{e^{i\frac{\Delta t}{2}}}{\sqrt{\mu }}\sin \sqrt{\mu }%
t\alpha _{1},  \nonumber \\
u_{ac}(t) &=&-i\frac{e^{i\frac{\Delta t}{2}}}{\sqrt{\mu }}\sin \sqrt{\mu }%
t\alpha _{2},  \nonumber \\
u_{ba}(t) &=&-i\alpha _{1}^{\dagger }\frac{e^{-i\frac{\Delta t}{2}}}{\sqrt{%
\mu }}\sin \sqrt{\mu }t,  \nonumber \\
u_{bb}(t) &=&1+\alpha _{1}^{\dagger }\alpha _{1}\frac{1}{\sqrt{\nu }}\frac{1%
}{\alpha _{1}^{\dagger }\alpha _{1}+\alpha _{2}^{\dagger }\alpha _{2}}\left[
e^{-i\frac{\Delta }{2}t}\left( i\frac{\Delta }{2}\sin \sqrt{\nu }t+\sqrt{\nu 
}\cos \sqrt{\nu }t\right) -\sqrt{\nu }\right] ,  \nonumber \\
u_{bc}(t) &=&\alpha _{1}^{\dagger }\alpha _{2}\frac{1}{\alpha _{1}^{\dagger
}\alpha _{1}+\alpha _{2}^{\dagger }\alpha _{2}}\left[ e^{-i\frac{\Delta }{2}%
t}\left( i\frac{\Delta }{2\sqrt{\nu }}\sin \sqrt{\nu }t+\cos \sqrt{\nu }%
t\right) -1\right] ,  \nonumber \\
u_{ca}(t) &=&-i\alpha _{2}^{\dagger }\frac{e^{-i\frac{\Delta t}{2}}}{\sqrt{%
\mu }}\sin \sqrt{\mu }t,  \nonumber \\
u_{cb}(t) &=&\alpha _{2}^{\dagger }\alpha _{1}\frac{1}{\alpha _{1}^{\dagger
}\alpha _{1}+\alpha _{2}^{\dagger }\alpha _{2}}\left[ e^{-i\frac{\Delta }{2}%
t}\left( i\frac{\Delta }{2\sqrt{\nu }}\sin \sqrt{\nu }t+\cos \sqrt{\nu }%
t\right) -1\right] ,  \nonumber \\
u_{cc}(t) &=&1+\alpha _{2}^{\dagger }\alpha _{2}\frac{1}{\sqrt{\nu }}\frac{1%
}{\alpha _{1}^{\dagger }\alpha _{1}+\alpha _{2}^{\dagger }\alpha _{2}}\left[
e^{-i\frac{\Delta }{2}t}\left( i\frac{\Delta }{2}\sin \sqrt{\nu }t+\sqrt{\nu 
}\cos \sqrt{\nu }t\right) -\sqrt{\nu }\right] ,
\end{eqnarray}%
where 
\begin{eqnarray}
\mu &=&\frac{\Delta ^{2}}{4}+\alpha _{1}\alpha _{1}^{\dagger }+\alpha
_{2}\alpha _{2}^{\dagger },  \nonumber \\
\nu &=&\frac{\Delta ^{2}}{4}+\alpha _{1}^{\dagger }\alpha _{1}+\alpha
_{2}^{\dagger }\alpha _{2}.
\end{eqnarray}

It is easy to show that for the non-degenerate case, i.e., 
\begin{eqnarray}
\alpha _{1} &=&g_{1}a_{1},  \nonumber \\
\alpha _{2} &=&g_{2}a_{2},
\end{eqnarray}%
we obtain 
\begin{eqnarray}
u_{aa}(t) &=&\frac{e^{i\frac{\Delta t}{2}}}{\sqrt{\mu }}\left[ \sqrt{\mu }%
\cos \sqrt{\mu }t-i\frac{\Delta }{2}\sin \sqrt{\mu }t\right] ,  \nonumber \\
u_{ab}(t) &=&-i\frac{e^{i\frac{\Delta t}{2}}}{\sqrt{\mu }}\sin \sqrt{\mu }%
t\alpha _{1},  \nonumber \\
u_{ac}(t) &=&-i\frac{e^{i\frac{\Delta t}{2}}}{\sqrt{\mu }}\sin \sqrt{\mu }%
t\alpha _{2},  \nonumber \\
u_{ba}(t) &=&-i\alpha _{1}^{\dagger }\frac{e^{-i\frac{\Delta t}{2}}}{\sqrt{%
\mu }}\sin \sqrt{\mu }t,  \nonumber \\
u_{bb}(t) &=&1+\alpha _{1}^{\dagger }\alpha _{1}\frac{1}{\sqrt{\nu _{1}}}%
\frac{1}{\alpha _{1}^{\dagger }\alpha _{1}+\alpha _{2}\alpha _{2}^{\dagger }}%
\left[ e^{-i\frac{\Delta }{2}t}\left( i\frac{\Delta }{2}\sin \sqrt{\nu _{1}}%
t+\sqrt{\nu _{1}}\cos \sqrt{\nu _{1}}t\right) -\sqrt{\nu _{1}}\right] , 
\nonumber \\
u_{bc}(t) &=&\alpha _{1}^{\dagger }\alpha _{2}\frac{1}{\alpha _{1}\alpha
_{1}^{\dagger }+\alpha _{2}^{\dagger }\alpha _{2}}\left[ e^{-i\frac{\Delta }{%
2}t}\left( i\frac{\Delta }{2\sqrt{\nu _{1}}}\sin \sqrt{\nu _{1}}t+\cos \sqrt{%
\nu _{1}}t\right) -1\right] ,  \nonumber \\
u_{ca}(t) &=&-i\alpha _{2}^{\dagger }\frac{e^{-i\frac{\Delta t}{2}}}{\sqrt{%
\mu }}\sin \sqrt{\mu }t,  \nonumber \\
u_{cb}(t) &=&\alpha _{2}^{\dagger }\alpha _{1}\frac{1}{\alpha _{1}^{\dagger
}\alpha _{1}+\alpha _{2}\alpha _{2}^{\dagger }}\left[ e^{-i\frac{\Delta }{2}%
t}\left( i\frac{\Delta }{2\sqrt{\nu _{1}}}\sin \sqrt{\nu _{1}}t+\cos \sqrt{%
\nu _{1}}t\right) -1\right] ,  \nonumber \\
u_{cc}(t) &=&1+\alpha _{2}^{\dagger }\alpha _{2}\frac{1}{\sqrt{\nu _{2}}}%
\frac{1}{\alpha _{1}\alpha _{1}^{\dagger }+\alpha _{2}^{\dagger }\alpha _{2}}%
\left[ e^{-i\frac{\Delta }{2}t}\left( i\frac{\Delta }{2}\sin \sqrt{\nu _{2}}%
t+\sqrt{\nu _{2}}\cos \sqrt{\nu _{2}}t\right) -\sqrt{\nu _{2}}\right] ,
\label{nondeg}
\end{eqnarray}%
where 
\begin{eqnarray}
\mu &=&\frac{\Delta ^{2}}{4}+\alpha _{1}\alpha _{1}^{\dagger }+\alpha
_{2}\alpha _{2}^{\dagger },  \nonumber \\
\nu _{1} &=&\frac{\Delta ^{2}}{4}+\alpha _{1}^{\dagger }\alpha _{1}+\alpha
_{2}\alpha _{2}^{\dagger },  \nonumber \\
\nu _{2} &=&\frac{\Delta ^{2}}{4}+\alpha _{1}\alpha _{1}^{\dagger }+\alpha
_{2}^{\dagger }\alpha _{2}.
\end{eqnarray}

Returning to the degenerate case, in the large detuning limit, we have 
\begin{eqnarray}
\sqrt{\mu } &\approx &\frac{\Delta ^{2}}{2}+\frac{\alpha _{1}\alpha
_{1}^{\dagger }+\alpha _{2}\alpha _{2}^{\dagger }}{\Delta }=\frac{\Delta }{2}%
+\frac{\mid g_{1}\mid ^{2}+\mid g_{2}\mid ^{2}}{\Delta }aa^{\dagger }, 
\nonumber \\
\sqrt{\nu } &\approx &\frac{\Delta ^{2}}{2}+\frac{\alpha _{1}^{\dagger
}\alpha _{1}+\alpha _{2}^{\dagger }\alpha _{2}}{\Delta }=\frac{\Delta }{2}+%
\frac{\mid g_{1}\mid ^{2}+\mid g_{2}\mid ^{2}}{\Delta }a^{\dagger }a,
\end{eqnarray}%
and we get 
\begin{eqnarray}
u_{aa}(t) &=&\exp \left( i\frac{\mid g_{1}\mid ^{2}+\mid g_{2}\mid ^{2}}{%
\Delta }taa^{\dagger }\right) ,  \nonumber \\
u_{ab}(t) &=&0,  \nonumber \\
u_{ac}(t) &=&0,  \nonumber \\
u_{ba}(t) &=&0,  \nonumber \\
u_{bb}(t) &=&1+\frac{\mid g_{1}\mid ^{2}}{\mid g_{1}\mid ^{2}+\mid g_{2}\mid
^{2}}\left[ \exp \left( i\frac{\mid g_{1}\mid ^{2}+\mid g_{2}\mid ^{2}}{%
\Delta }ta^{\dagger }a\right) -1\right] ,  \nonumber \\
u_{bc}(t) &=&\frac{g_{1}^{\ast }g_{2}}{\mid g_{1}\mid ^{2}+\mid g_{2}\mid
^{2}}\left[ \exp \left( i\frac{\mid g_{1}\mid ^{2}+\mid g_{2}\mid ^{2}}{%
\Delta }ta^{\dagger }a\right) -1\right] ,  \nonumber \\
u_{ca}(t) &=&0,  \nonumber \\
u_{cb}(t) &=&\frac{g_{2}^{\ast }g_{1}}{\mid g_{1}\mid ^{2}+\mid g_{2}\mid
^{2}}\left[ \exp \left( i\frac{\mid g_{1}\mid ^{2}+\mid g_{2}\mid ^{2}}{%
\Delta }ta^{\dagger }a\right) -1\right] ,  \nonumber \\
u_{cc}(t) &=&1+\frac{\mid g_{2}\mid ^{2}}{\mid g_{1}\mid ^{2}+\mid g_{2}\mid
^{2}}\left[ \exp \left( i\frac{\mid g_{1}\mid ^{2}+\mid g_{2}\mid ^{2}}{%
\Delta }ta^{\dagger }a\right) -1\right] .
\end{eqnarray}%
If 
\begin{eqnarray}
g_{1} &=&ge^{i\varphi _{1}},  \nonumber \\
g_{2} &=&ge^{i\varphi _{2}},
\end{eqnarray}%
we finally have 
\begin{eqnarray}
u_{aa}(t) &=&\exp \left( i\frac{2g^{2}t}{\Delta }\right) \exp \left( i\frac{%
2g^{2}t}{\Delta }a^{\dagger }a\right) ,  \nonumber \\
u_{ab}(t) &=&0,  \nonumber \\
u_{ac}(t) &=&0,  \nonumber \\
u_{ba}(t) &=&0,  \nonumber \\
u_{bb}(t) &=&\frac{1}{2}\left[ \exp \left( i\frac{2g^{2}t}{\Delta }%
a^{\dagger }a\right) +1\right] ,  \nonumber \\
u_{bc}(t) &=&\frac{e^{i(\varphi _{1}-\varphi _{2})}}{2}\left[ \exp \left( i%
\frac{2g^{2}t}{\Delta }a^{\dagger }a\right) -1\right] ,  \nonumber \\
u_{ca}(t) &=&0,  \nonumber \\
u_{cb}(t) &=&\frac{e^{-i(\varphi _{1}-\varphi _{2})}}{2}\left[ \exp \left( i%
\frac{2g^{2}t}{\Delta }a^{\dagger }a\right) -1\right] ,  \nonumber \\
u_{cc}(t) &=&\frac{1}{2}\left[ \exp \left( i\frac{2g^{2}t}{\Delta }%
a^{\dagger }a\right) +1\right] ,
\end{eqnarray}%
which agrees with the result obtained in \cite{Knight}.

Now, let us considering the classical limit in the three level lambda atom
interacting with two dephased modes with the same frequency (see (\ref%
{nondeg})). We see that the matrix elements of the evolution operator read
as:

\begin{eqnarray}
u_{bb} &=&1+\epsilon _{1}^{\ast }\epsilon _{1}\frac{1}{2\mid \epsilon \mid
^{2}}[e^{-i\Delta t/2}(\cos \sqrt{2\mid \epsilon \mid ^{2}+\Delta ^{2}/4}t 
\nonumber \\
&&+i\frac{\Delta }{2\sqrt{2\mid \epsilon \mid ^{2}+\Delta ^{2}/4}}\sin \sqrt{%
2\mid \epsilon h\mid ^{2}+\Delta ^{2}/4}t)-1],  \nonumber \\
u_{cc} &=&1+\epsilon _{2}^{\ast }\epsilon _{2}\frac{1}{2\mid \epsilon \mid
^{2}}[e^{-i\Delta t/2}(\cos \sqrt{2\mid \epsilon \mid ^{2}+\Delta ^{2}/4}t 
\nonumber \\
&&+i\frac{\Delta }{2\sqrt{2\mid \epsilon \mid ^{2}+\Delta ^{2}/4}}\sin \sqrt{%
2\mid \epsilon \mid ^{2}+\Delta ^{2}/4}t)-1],  \nonumber \\
u_{cb} &=&\epsilon _{1}^{\ast }\epsilon _{2}\frac{1}{2\mid \epsilon \mid ^{2}%
}[e^{-i\Delta t/2}(\cos \sqrt{2\mid \epsilon \mid ^{2}+\Delta ^{2}/4}t 
\nonumber \\
&&+i\frac{\Delta }{2\sqrt{2\mid \epsilon \mid ^{2}+\Delta ^{2}/4}}\sin \sqrt{%
2\mid \epsilon \mid ^{2}+\Delta ^{2}/4}t)-1],  \nonumber \\
u_{bc} &=&\epsilon _{1}\epsilon _{2}^{\ast }\frac{1}{2\mid \epsilon \mid ^{2}%
}[e^{-i\Delta t/2}(\cos \sqrt{2\mid \epsilon \mid ^{2}+\Delta ^{2}/4}t 
\nonumber \\
&&+i\frac{\Delta }{2\sqrt{2\mid \epsilon \mid ^{2}+\Delta ^{2}/4}}\sin \sqrt{%
2\mid \epsilon \mid ^{2}+\Delta ^{2}/4}t)-1],
\end{eqnarray}%
where we have used 
\begin{eqnarray}
g_{1}a_{1} &\rightarrow &\epsilon _{1}=\epsilon e^{i\theta _{1}},  \nonumber
\\
g_{2}a_{2} &\rightarrow &\epsilon _{2}=\epsilon e^{i\theta _{2}}.
\end{eqnarray}%
In the high detuning limit we obtain the expression 
\begin{eqnarray}
u_{bb} &=&\frac{1}{2}(e^{i\varphi }+1),  \nonumber \\
u_{cc} &=&\frac{1}{2}(e^{i\varphi }+1),  \nonumber \\
u_{cb} &=&\frac{1}{2}e^{-i\phi }(e^{i\varphi }-1),  \nonumber \\
u_{bc} &=&\frac{1}{2}e^{i\phi }(e^{i\varphi }-1),
\end{eqnarray}%
where $\phi =\theta _{1}-\theta _{2}$ and $\varphi =2\mid \epsilon \mid
^{2}t/\Delta $ . Choosing $\varphi =\pi /2$ and $\phi =\pi /2$ we get 
\begin{equation}
U=\frac{1}{\sqrt{2}}e^{i\pi /4}\left( 
\begin{array}{ll}
1 & -1 \\ 
1 & 1%
\end{array}%
\right) .
\end{equation}%
Choosing $\varphi =\pi $ and $\phi =\pi /2$ we get 
\begin{equation}
U=i\left( 
\begin{array}{ll}
0 & -1 \\ 
1 & 0%
\end{array}%
\right) .  \label{Uc1}
\end{equation}%
Choosing $\varphi =\pi $ and $\phi =-\pi /2$ we get 
\begin{equation}
U=i\left( 
\begin{array}{ll}
0 & 1 \\ 
-1 & 0%
\end{array}%
\right) .
\end{equation}%
Choosing $\varphi =\pi $ and $\phi =\pi $ we get 
\begin{equation}
U=\left( 
\begin{array}{ll}
0 & 1 \\ 
1 & 0%
\end{array}%
\right) .
\end{equation}

\textbf{Figure Captions} \newline

\textbf{Fig. 1-} Energy states scheme of a three-level atom where $|e\rangle 
$ is the upper state with atomic frequency $\omega _{e}$, $\ |f\rangle $ is
the intermediate state with atomic frequency $\omega _{f}$, $|g\rangle $ is
the lower state with atomic frequency $\omega _{g}$ and $\omega $ is the
cavity field frequency and $\Delta =(\omega _{e}-\omega _{f})-\omega $ is
the detuning. The transition $\mid f\rangle \rightleftharpoons \mid e\rangle 
$ is far enough of resonance with the cavity central frequency such that
only virtual transitions occur between these levels (only these states
interact with field in cavity $C$). In addition we assume that the
transition $\mid e\rangle \rightleftharpoons \mid g\rangle $ is highly
detuned from the cavity frequency so that there will be no coupling with the
cavity field in $C$.\newline

\textbf{Fig. 2- } Set-up for the GHZ experiment. Atom $A0$ is sent to the
cavity $R0,$ where it is prepared in a coherent superposition, and to cavity 
$C$ to prepare it in a state which is a superposition of zero and one Fock
states. Atom $A1$ passes through \ the Ramsey cavity $R1$ where it is
prepared in a coherent superposition, cavity $C$ and through the Ramsey
cavity $R2.$ Atom $A2$ passes through \ the Ramsey cavity $R3$ where it is
prepared in a coherent superposition, cavity $C$ and through the Ramsey
cavity $R4$. Atom $A3$ is prepared in state $|g_{3}\rangle $, and sent to
the cavity $C$. Once the GHZ state has been obtained, the GHZ test is
performed making use of the Ramsey cavities $K1,K2$ and $K3$ and detectors $%
D1$, $D2$ and $D3$ as described in the text.\newline

\textbf{Fig. 3-} Energy level scheme of the three-level lambda atom where $%
|a\rangle $ is the upper state with atomic frequency $\omega _{a}$, $%
|b\rangle $ \ and $|c\rangle $ are the lower states with atomic frequency $%
\omega _{b}$ and $\omega _{c}$, $\omega $ is the cavity field frequency and $%
\Delta =\omega _{a}-\omega _{b}-\omega =\omega _{a}-\omega _{c}-\omega $ is
the detuning.\newline

\textbf{Fig. 4- }Set-up for the GHZ experiment. Atom $A0$ is sent to the
cavity $R0,$ where it is prepared in a coherent superposition, and to cavity 
$C$ to prepare it in a state which is a superposition of zero and one Fock
states. Atom $A1$ is prepared in state $|b_{1}\rangle $ and passes through \
cavity $C.$ Atom $A2$ is prepared in state $|b_{2}\rangle $ and passes
through cavity $C$. Atom $A3$ is prepared in state $|g_{3}\rangle $, and
sent to the cavity $C$. Once the GHZ state has been obtained, the GHZ test
is performed making use of the Ramsey cavities $K1,K2$ and $K3$ and
detectors $D1$, $D2$ and $D3$ as described in the text.

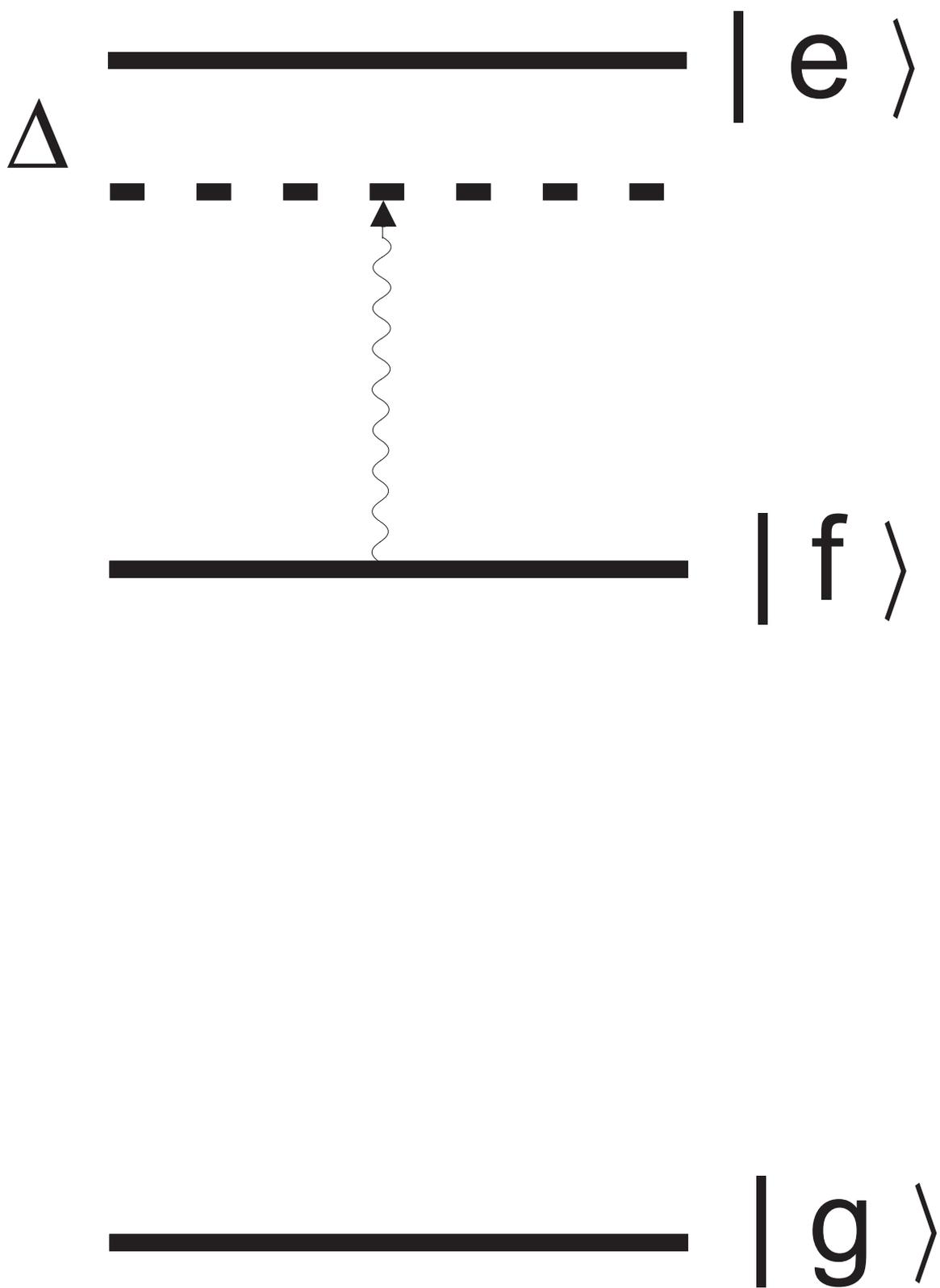
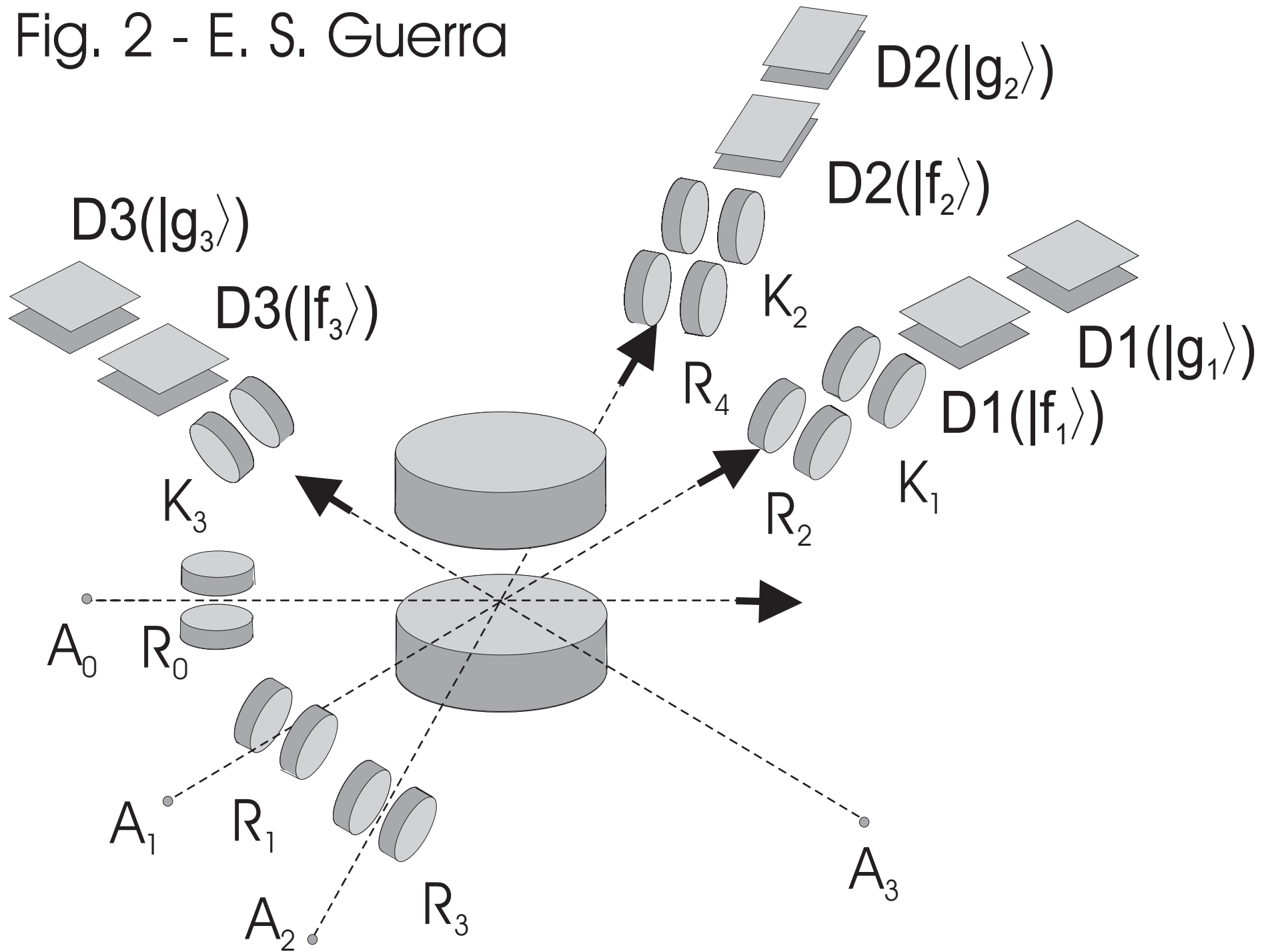
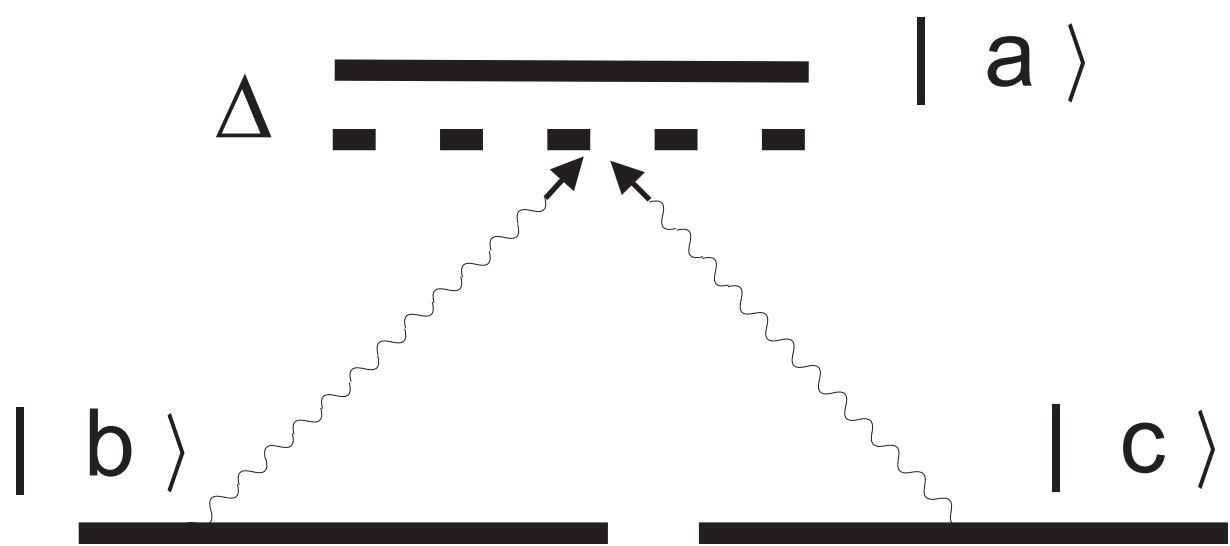
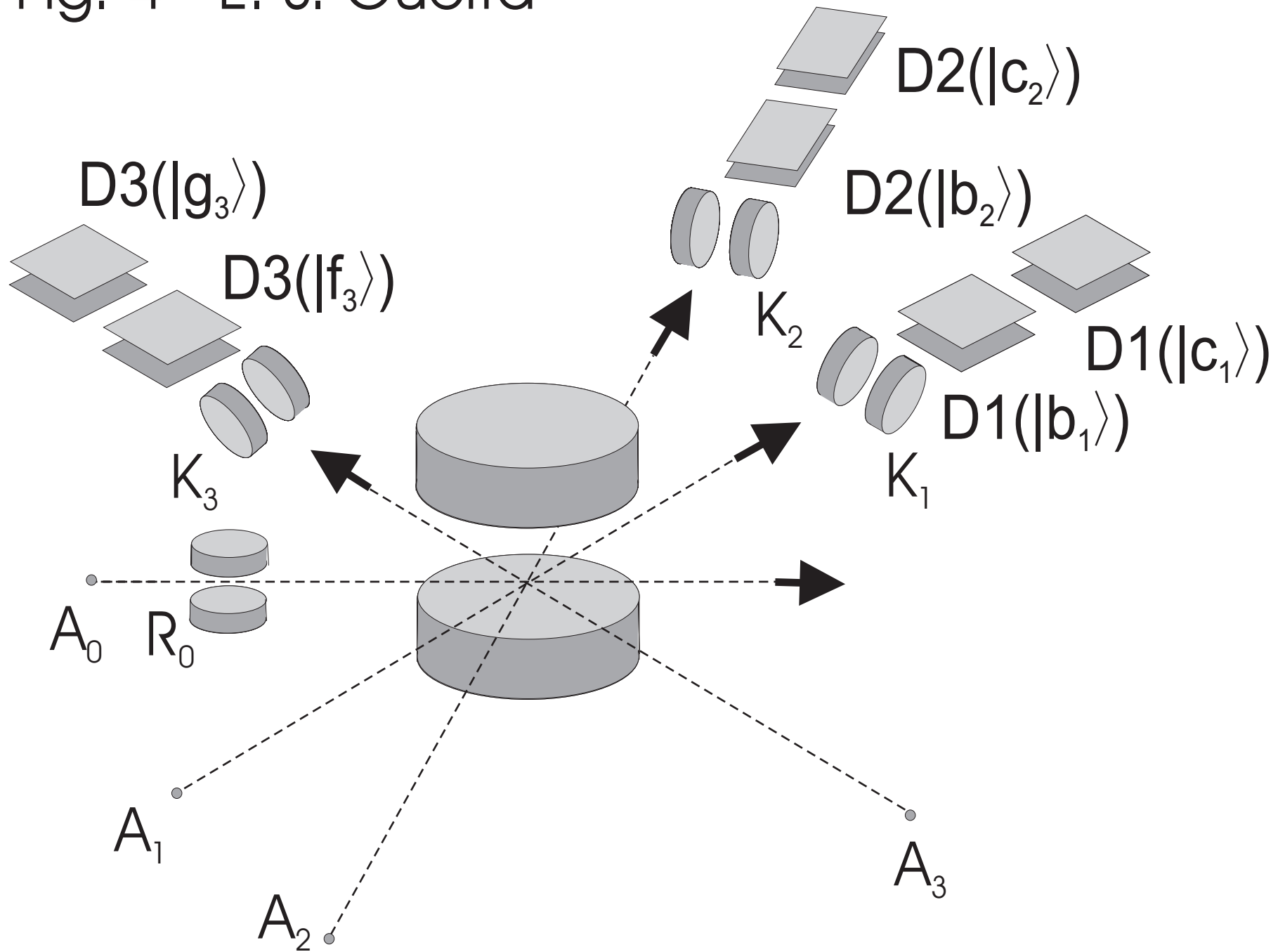
\end{document}